\documentclass{aa}
\usepackage{amssymb}
\usepackage{graphicx}
\usepackage{txfonts}

\newcommand{\arcmins}{\mbox{$^{\prime}$}}
\newcommand{\parcsec}{\mbox{$\stackrel{\prime\prime}{\textstyle .}$}}

\newcommand{\psec}{\mbox{$\stackrel{s}{\textstyle .}$}}
\newcommand{\hours}{\mbox{$^{h}$}}
\newcommand{\mins}{\mbox{$^{m}$}}

\begin{document}
\title{The line-of-sight towards GRB 030429 at $z=2.66$:\\
Probing the matter at stellar, galactic and intergalactic
scales\thanks{Based on 
  observations made with ESO
  Telescopes at the Paranal Observatory by GRACE under programme 
  ID 71.D-0355(A+B+C).}}

\author{P.~Jakobsson
       \inst{1,2} \and
       J.~Hjorth
       \inst{1} \and
       J.~P.~U.~Fynbo
       \inst{1,3} \and
       M.~Weidinger
       \inst{3,4} \and
       J.~Gorosabel
       \inst{5,6} \and
       C.~Ledoux
       \inst{7} \and
       D.~Watson
       \inst{1} \and
       G.~Bj\"ornsson
       \inst{2} \and
       E.~H.~Gudmundsson
       \inst{2} \and
       R.~A.~M.~J.~Wijers
       \inst{8} \and
       P.~M\o ller
       \inst{4} \and
       K.~Pedersen
       \inst{1} \and
       J.~Sollerman
       \inst{9} \and
       A.~A.~Henden
       \inst{10} \and
       B.~L.~Jensen
       \inst{1} \and
       A.~Gilmore
       \inst{11} \and
       P.~Kilmartin
       \inst{11} \and
       A.~Levan
       \inst{12} \and
       J.~M.~Castro Cer\'{o}n
       \inst{6} \and
       A.~J.~Castro-Tirado
       \inst{5} \and
       A.~Fruchter
       \inst{6} \and
       C.~Kouveliotou
       \inst{13} \and
       N.~Masetti 
       \inst{14} \and
        N.~Tanvir
       \inst{15}
}

\offprints{P. Jakobsson, \\ \email{pallja@astro.ku.dk}}
   
\institute{Niels Bohr Institute, Astronomical Observatory,
     University of Copenhagen, Juliane Maries Vej 30, DK-2100 
     Copenhagen, Denmark
\and
     Science Institute, University of Iceland, Dunhaga 3,
     107 Reykjav\'{\i}k, Iceland
\and
     Department of Physics and Astronomy, University of Aarhus, 
     Ny Munkegade, DK-8000 \AA rhus C, Denmark
\and
     European Southern Observatory, Karl-Schwarzschild-Stra\ss e 2, 85748, 
     Garching bei M\"unchen, Germany
\and
     Instituto de Astrof\'{\i}sica de Andaluc\'{\i}a (IAA-CSIC),
     P.O. Box 03004, E-18080 Granada, Spain
\and
     Space Telescope Science Institute, 3700 San Martin
     Drive, Baltimore, MD 21218, USA
\and
     European Southern Observatory, Alonso de C\'ordova 3107, 
     Casilla 19001, Santiago 19, Chile
\and
     Astronomical Institute, University of Amsterdam,
     Kruislaan 403, 1098 SJ Amsterdam, The Netherlands      
\and
     Stockholm Observatory, Department of Astronomy, AlbaNova, 106 91 
     Stockholm, Sweden
\and
     USRA/USNO Flagstaff Station, P. O. Box 1149, Flagstaff, 
     AZ 86002 USA
\and
     Mt John Observatory, Department of Physics and Astronomy,
     University of Canterbury, Christchurch 8020, New Zealand 
\and
     Department of Physics and Astronomy, University of Leicester,
     University Road, Leicester, LE1 7RH, UK
\and
     NASA MSFC, SD-50 Huntsville, AL 35812, USA
\and
     Istituto di Astrofisica Spaziale e Fisica Cosmica, Sezione di 
     Bologna, CNR, Via Gobetti 101, 40129 Bologna, Italy
\and
     Department of Physical Sciences, University of Hertfordshire,
     College Lane, Hatfield, Herts AL10 9AB, UK 
}
   \date{Received 5 May 2004 / Accepted ???}

\abstract{We report the discovery of the optical afterglow (OA) of the
  long-duration gamma-ray burst GRB~030429, and present a 
  comprehensive optical/near-infrared dataset used to probe the matter
  at different distance scales, i.e. in the burst environment, in the host
  galaxy and in an intervening absorber. A break in the afterglow light 
  curve is seen approximately 1\,day from the onset of the burst. 
  The light
  curve displays a significant deviation from a simple broken power-law 
  with a bright 1.5 magnitude bump with a duration of 2--3\,days.
  The optical/near-infrared spectral energy distribution is best fit
  with a power-law with index $\beta = -0.36 \pm 0.12$ reddened by an
  SMC-like extinction law with (a modest) $A_V = 0.34 \pm 0.04$. 
  In addition, we present deep spectroscopic observations obtained with the 
  Very Large Telescope. The redshift measured via metal absorption lines 
  in the OA is $z = 2.658 \pm 0.004$. Based on the damped Ly$\alpha$ 
  absorption line in the OA spectrum we measure the \ion{H}{i} column 
  density to be $\log N(\ion{H}{i}) = 21.6 \pm 0.2$. This confirms
  the trend that GRBs tend to be located behind very large \ion{H}{i}
  column densities. The resulting dust-to-gas ratio is 
  consistent with that found in the SMC, indicating a low
  metallicity and/or a low dust-to-metal ratio in the burst  
  environment. We find that a neighbouring galaxy, at a separation 
  of only $1\parcsec2$, has $z = 0.841 \pm 0.001$, ruling it out as 
  the host of GRB~030429. The small impact parameter of this 
  nearby galaxy, which is responsible for \ion{Mg}{ii} absorption in
  the OA spectrum, is in contrast
  to previous identifications of most QSO absorption-selected galaxy 
  counterparts. Finally, we demonstrate that the OA was not affected 
  by strong gravitational lensing via the nearby galaxy.

}

\titlerunning{The line-of-sight towards GRB~030429 at $z=2.66$}

\authorrunning{P. Jakobsson et al.}

\maketitle
\section{Introduction}
\begin{figure}
   \centering
   \resizebox{\hsize}{!}{\includegraphics[bb=90 374 540 700,clip]{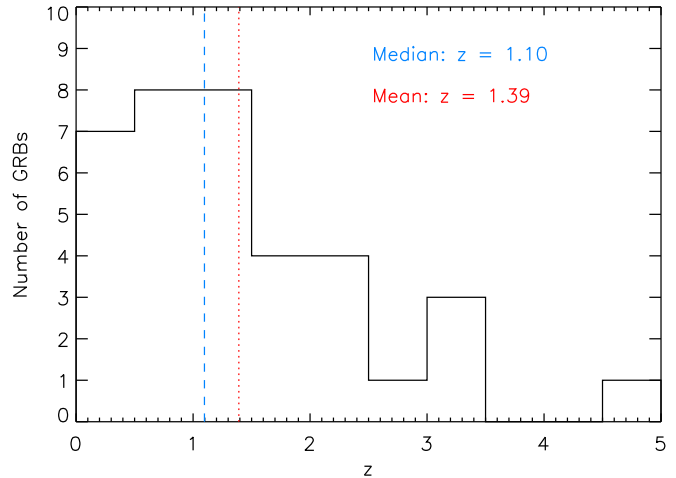}}
      \caption{A histogram showing the distribution of GRB redshifts
   known to date (July 2004). The median value of the 36 redshifts 
   is indicated by the dashed line, while the mean value is 
   shown with the dotted line.}
         \label{z.fig}
   \end{figure}

Considerable progress has been made in the understanding of gamma-ray 
bursts (GRBs) since BeppoSAX started distributing localisations with
arcmin precision and van Paradijs et al. (\cite{paradijs}) discovered 
the first optical afterglow (OA). Of the approximately 50 OAs that have 
been detected since 1997, 36 have had their redshift spectroscopically 
determined. As shown in Fig.~\ref{z.fig}, the mean redshift is around 
1.4 with 13 GRBs at $z > 1.5$.
\par
Independently of the brightness of GRB host galaxies, OAs can be used as
tools to obtain information about the gas, metallicity and dust content
of the host. In particular, the optical/near-infrared (NIR) 
spectral energy distribution 
(SED) can provide the extinction, while the neutral hydrogen column 
density, $N(\ion{H}{i})$, can be derived from the damped Ly$\alpha$ 
absorption spectroscopy, provided that the burst is distant enough 
for the Ly$\alpha$
line to be redshifted into the optical/near-UV domain. The detection 
of damped Ly$\alpha$ absorption lines in the spectra of several GRB
afterglows (Jensen et al. \cite{brian}; Fynbo et al 
\cite{damped}; Hjorth et al. \cite{jens}; Vreeswijk et al. 
\cite{paul}) is consistent with the now firmly established link 
between long-duration GRBs and core-collapse of massive stars (e.g. 
Hjorth et al. \cite{jens_nat}; Stanek et al. \cite{stanek}).
\par 
Like QSOs, GRB afterglows are distant sources that can be used
as cosmological beacons to investigate intervening galaxies unrelated 
to the burst itself. For the current discussion, the most important 
difference between these two astronomical phenomena is that 
OAs are ephemeral, vanishing altogether 
within a couple of months. This leaves the line-of-sight clear and
without any interference from a bright object. Another significant 
difference is that afterglows typically have featureless spectra, 
making it easier to identify intervening absorption systems.
\par
In this paper we present the detection of the OA of GRB~030429 and 
use it to probe the progenitor environment (sub-pc scale), the host 
galaxy properties (kpc scale) and an intervening \ion{Mg}{ii} 
absorber (Gpc scale). The organisation of the paper is as follows. 
The optical, NIR and spectroscopic observations are 
presented in Sect.~\ref{obs.sec}. In Sect.~\ref{ima.sec} we investigate 
the optical/NIR light curve and the SED.
We analyze the spectrum and give the redshift of GRB~030429 in 
Sect.~\ref{spec.sec}. In addition we fit a \ion{H}{i} column 
density model to a damped Ly$\alpha$ line present in the OA spectrum.
In Sect.~\ref{dis.sec} we use the derived properties of the OA to 
compare our results with afterglow models. We also discuss the
properties of the nearby \ion{Mg}{ii} absorbing galaxy and assess whether 
it could be responsible for strong gravitational lensing of the OA. 
Finally, the main results are summarised in Sect.~\ref{con.sec}. 
\par
We adopt a cosmology where the Hubble para\-meter is 
$H_0 = 70$\,km\,s$^{-1}$\,Mpc$^{-1}$,
$\Omega_{\mathrm{m}} = 0.3$ and $\Omega_{\Lambda} = 0.7$. For these
parameters, a redshift of 2.66 (0.84) corresponds to a luminosity distance
of 21.99\,Gpc (5.33\,Gpc) and a distance modulus of 46.7 (43.6). One 
arcsecond is equivalent to 7.96 (7.63) proper kiloparsecs, and the 
lookback time is 11.1\,Gyr (7.0\,Gyr).
\section{Observations}
\label{obs.sec}
GRB 030429 was detected by the French Gamma Telescope 
(FREGATE), Wide Field X-ray Monitor (WXM) and Soft X-ray Camera (SXC) 
on-board the HETE-2 satellite on 2003 April 29.446 
UT. The burst had a duration of 10.3\,s, placing it in the 
``long-duration'' burst category. Initially, a $1\arcmin$ radius error 
circle was circulated via the GRB Coordinate Network
(GCN)\footnote{\texttt{http://gcn.gsfc.nasa.gov/gcn/}} approximately 
2\,hours after the burst. A week later the correct error radius
of $2\arcmin$ was distributed (Doty et al. \cite{doty}). 
\par
\begin{figure}
   \centering
   \resizebox{\hsize}{!}{\includegraphics{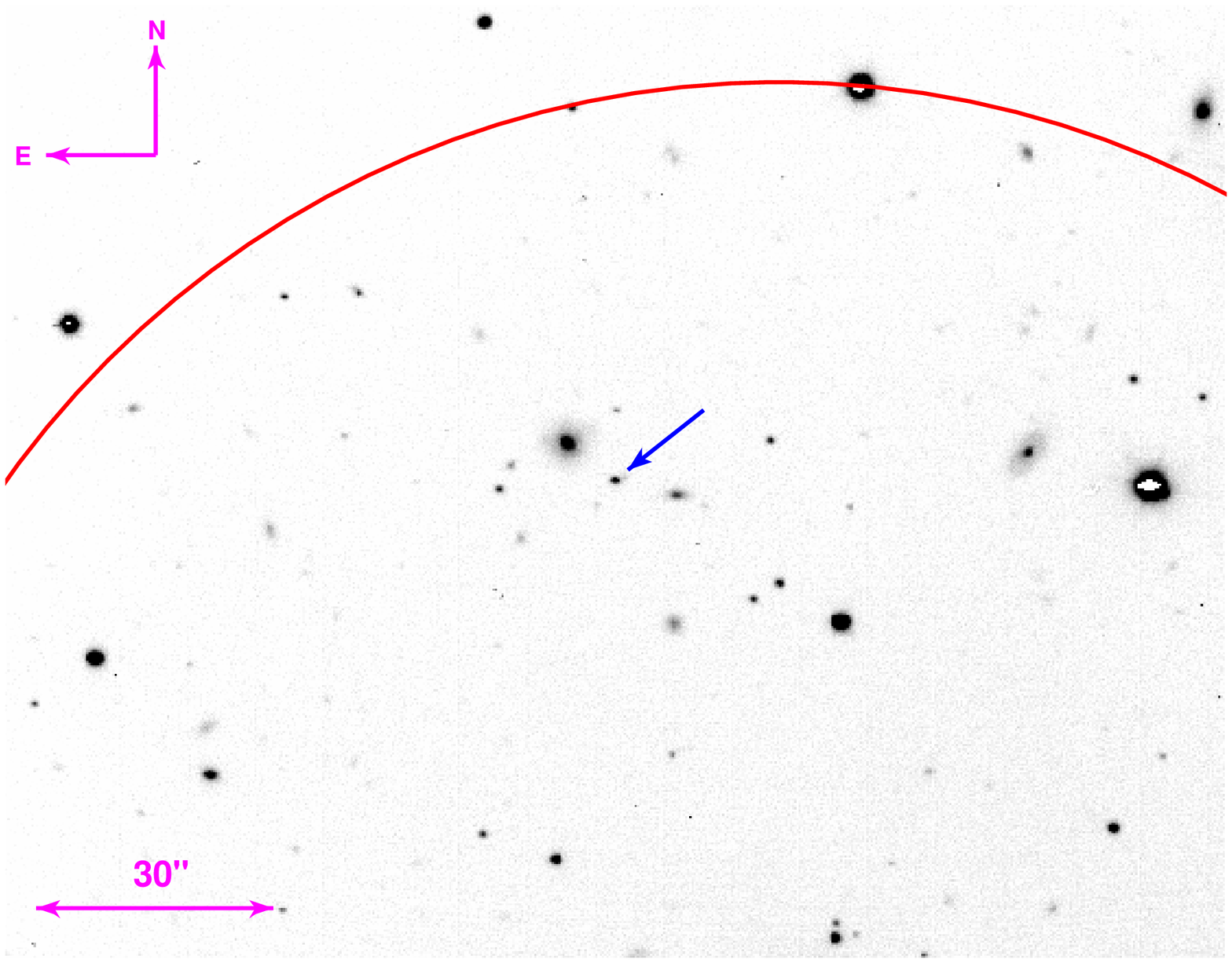}}
   \resizebox{\hsize}{!}{\includegraphics{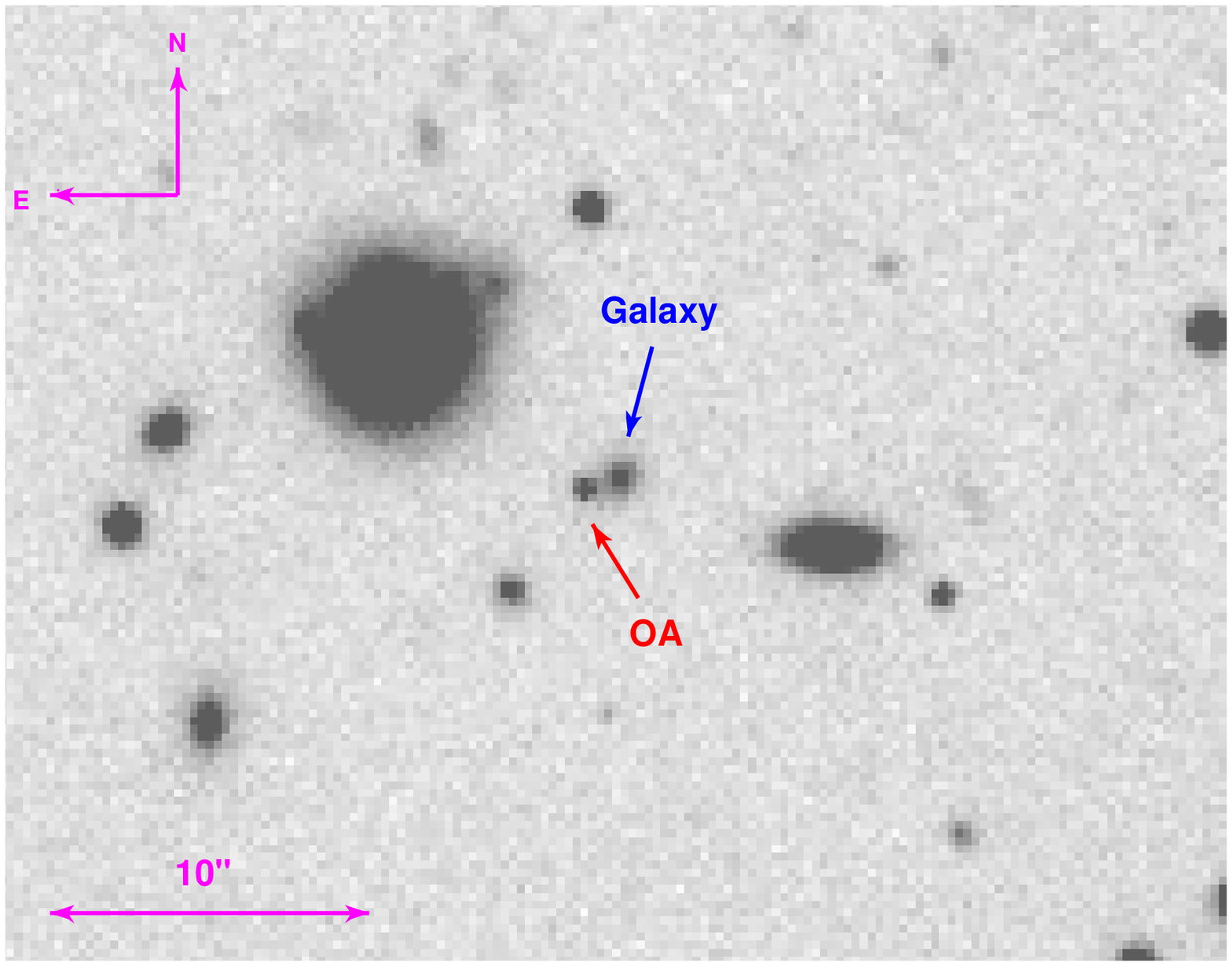}}
      \caption{\emph{Top:} a 180\,s $R$-band VLT/FORS2 image of the GRB~030429 
      optical afterglow, obtained $\sim$13\,hours after the burst occurred. 
      The position of the optical afterglow is marked with an arrow. The 
      SXC localization is shown with a continuous circle of radius $2\arcmin$. 
      \emph{Bottom:} a combined 900\,s $R$-band VLT/FORS2 image with a
      seeing of $0\parcsec5$, taken $\sim$3.6\,days 
      from the onset of the burst. The field of view is smaller 
      than shown in the top panel in order to emphasise the surroundings
      of the OA. A galaxy at a distance of $1\parcsec2$ from the OA is 
      clearly visible and marked with an arrow.}
         \label{ot.fig}
   \end{figure}

The OA was discovered at the Mt. John 0.6-m 
telescope $\sim$3.5\,hours after the burst (Gilmore et al. 
\cite{gilmore}). It was identified outside the 
initial erroneous SXC $1\arcmin$ \mbox{error circle} after comparison 
with a DSS-2 red plate. The OA was monitored in the optical during the 
following days with the \mbox{1.0-m} and \mbox{1.55-m} telescopes at 
the U.S. Naval Observatory Flagstaff Station (NOFS), and with the 
Very Large Telescope (VLT), using either the 
Antu/FORS1 or the Yepun/FORS2 combination (Fynbo et al. \cite{johanGCN}). 
In addition, Antu/ISAAC was used to obtain NIR images of the 
OA shortly after the burst. A VLT/FORS2 image of the OA and its 
surrounding field is displayed in Fig.~\ref{ot.fig}. The journal 
of our observations is given in Table~\ref{phot.tab}. 
\begin{table}[!t]
\caption[]{A log of the GRB~030429 follow-up imaging observations.
Upper limits are $2 \sigma$ in a circular aperture with 
radius $1\arcsec$. No correction for Galactic extinction has been 
applied to the photometry.}
\label{phot.tab}
\begin{minipage}{\columnwidth}
\centering
\setlength{\arrayrulewidth}{0.8pt}   
\begin{tabular}{rcccc@{}c}
\hline
\hline
\vspace{-2 mm} \\
$\Delta t$\footnote{Days after 2003 April 29.446} \hspace{0.8 mm} 
& Telescope/ & 
Magnitude\footnote{The 0.6-m observations were unfiltered but tied to
the $R$-band as described in the text.} & Seeing  
& Exp. time \\
  \hspace{-0 mm}[days]    &   Instrument    &       & [arcsec] & [s] \\
\vspace{-2 mm} \\
\hline
\vspace{-2 mm} \\
\hspace{-3 mm}
\emph{B-band:}      &            &                    &      & \\    
1.747 & 1.0-m NOFS     & $22.24 \pm 0.30$ & 2.3 & $720$         \\
\hspace{-3 mm}
\emph{V-band:}    &            &                    &      & \\    
0.535 & Yepun/FORS2    & $21.45 \pm 0.03$ & 0.7 & $5 \times 180$ \\
0.764 & Yepun/FORS2    & $21.79 \pm 0.03$ & 0.8 & $5 \times 180$ \\
1.755 & 1.0-m NOFS     & $21.89 \pm 0.32$ & 2.4 & $480$ \\
2.541 & Yepun/FORS2    & $23.18 \pm 0.11$ & 0.9 & $5 \times 180$ \\
6.631 & Yepun/FORS2    & $>$26.0          & 1.3 & $5 \times 180$ \\
\hspace{-3 mm}
\emph{R-band:}         &            &                    &      & \\    
0.145 & 0.6-m Mt. John & $19.67 \pm 0.11$ & 3.4 & $10 \times 60$ \\ 
0.154 & 0.6-m Mt. John & $19.39 \pm 0.08$ & 3.3 & $10 \times 60$ \\
0.170 & 0.6-m Mt. John & $19.62 \pm 0.09$ & 3.3 & $10 \times 60$ \\
0.214 & 0.6-m Mt. John & $19.56 \pm 0.11$ & 4.8 & $10 \times 60$ \\
0.548 & Yepun/FORS2    & $20.86 \pm 0.04$ & 0.7 & $5 \times 180$ \\
0.777 & Yepun/FORS2    & $21.13 \pm 0.04$ & 0.8 & $5 \times 180$ \\
1.761 & 1.0-m NOFS     & $21.42 \pm 0.27$ & 2.2 & $420$ \\
1.862 & 1.55-m NOFS    & $21.27 \pm 0.09$ & 1.2 & $600$ \\
1.884 & 1.55-m NOFS    & $21.58 \pm 0.15$ & 1.4 & $600$ \\
2.553 & Yepun/FORS2    & $22.54 \pm 0.06$ & 0.9 & $5 \times 180$ \\
2.793 & 1.55-m NOFS    &  $22.55 \pm 0.09$ & 1.3 & $5 \times 600$  \\
3.632 & Yepun/FORS2    & $23.71 \pm 0.12$ & 0.5 & $5 \times 180$ \\
6.644 & Yepun/FORS2    & $25.20 \pm 0.30$ & 1.2 & $5 \times 180$ \\
67.641 & Antu/FORS1    & $>$26.3          & 0.7 & $15 \times 180$ \\
\hspace{-3 mm}
\emph{I-band:}    &            &                    &      & \\
0.561 & Yepun/FORS2    & $20.29 \pm 0.05$ & 0.7 & $5 \times 180$ \\
0.790 & Yepun/FORS2    & $20.63 \pm 0.05$ & 0.8 & $5 \times 180$ \\ 
1.766 & 1.0-m NOFS     & $20.51 \pm 0.37$ & 2.2 & $420$          \\
2.566 & Yepun/FORS2    & $21.72 \pm 0.09$ & 0.6 & $5 \times 180$ \\ 
6.658 & Yepun/FORS2    & $24.70 \pm 0.30$ & 1.1 & $5 \times 180$ \\ 
\hspace{-3 mm}
\emph{J$_s$-band:}&            &                    &      & \\    
0.538 & Antu/ISAAC     & $19.25 \pm 0.04$ & 0.6 & $20 \times 90$ \\
0.753 & Antu/ISAAC     & $19.51 \pm 0.04$ & 0.7 & $20 \times 90$ \\
\hspace{-3 mm}
\emph{K$_s$-band:}&            &                    &      & \\  
0.568 & Antu/ISAAC     & $17.70 \pm 0.06 $ & 0.6 & $30 \times 60$ \\
0.783 & Antu/ISAAC     & $18.01 \pm 0.06 $ & 0.6 & $30 \times 60$ \\
\vspace{-2 mm} \\
\hline
\end{tabular}
\end{minipage}
\end{table}     
\par
Using Antu we obtained $11\times 600$\,s spectra of the OA starting on
2003 May 2.214 UT, 2.77\,days after the burst (Weidinger et al. 
\cite{weidA}, \cite{weidB}). The data were acquired with the FORS1 
instrument in long-slit spectroscopy mode with the
G300V grism, GG375 order separation filter and a $1\parcsec3$ wide slit. 
This resulted in a wavelength
coverage from 3900--8000\,\AA. The slit was oriented with a position angle
of $-71\fdg1$ in order to cover both the OA and a
neighbouring galaxy (lower panel of Fig.~\ref{ot.fig}). The observations 
were done with the standard
resolution collimator. The individual spectra were combined,
yielding a seeing of $0\parcsec8$ and a spectral resolution of
10\,\AA\ FWHM. The projected pixel size was $0\parcsec2 
\times 2.58$\,\AA.
\par
Berger \& 
Frail (\cite{berger}) observed the error circle of GRB~030429 
with the Very Large Array at 8.46~GHz, 2.65, 5.63 and 16.68\,days
after the burst. No radio sources were detected at the position of 
the OA down to a 3$\sigma$ limit of 0.18, 0.16 and 0.10\,mJy, 
respectively.
\section{Imaging}
\label{ima.sec}
\subsection{Photometry}
The data were reduced using standard techniques for de-biasing and 
flat-fielding. The photometry of the afterglow was carried out using 
point-spread function (PSF) fitting photometry (Stetson 
\cite{stetson, stetson2}). For the VLT images, the relative optical 
magnitudes were transformed to the Johnson photometric 
system using observations 
of Landolt (\cite{landolt}) standards. For the other telescopes we
used the calibrated magnitudes of stars in the field (Henden 
\cite{henden}). We note that the 0.6-m observations were unfiltered,
but we tied them to the $R$-band using Henden (\cite{henden}) standards
with similar colours as the OA. For the NIR data we used 2MASS stars 
in the field to transform our observations to the standard system. 
\par
As seen in Table~\ref{phot.tab} the afterglow was observed at
the first two VLT epochs ($\Delta t \approx 0.55$\,days and
\mbox{$\Delta t \approx 0.78$\,days}) in all five filters ($VRIJ_sK_s$) 
and in $VRI$ during the third VLT epoch ($\Delta t 
\approx 2.55$\,days). The final detection of the OA is 
in $R$ and $I$ at $\Delta t \approx 6.65$\,days, where 
$\Delta t$ is the time from the onset of the burst. We also have 
an upper limit in $V$ at the same time, and a deep 
upper limit in $R$ roughly 2~months after the burst. We note that 
there is a galaxy $\sim$$1\parcsec2$ WNW of the OA, visible in the 
lower panel of Fig.~\ref{ot.fig}, which was initially considered to
be the probable host galaxy (see Sect.~\ref{redshift.sec}). 
There is no indication of a host 
directly underlying the OA; we can set an upper limit of 
$R > 26.3$ (2$\sigma$) from our last observation. For a future host 
search we include the accurate position of the afterglow based on USNO 
CCD Astrograph Catalog (UCAC2) 
stars in the field: RA(J2000) = $12\hours13\mins07\psec495(10)$ and
Dec(J2000) = $-20^{\circ}54\arcmins49\parcsec54(3)$
with the error in the last digits indicated in parentheses. The 
accuracy of the afterglow position in RA is worse than in 
declination due to the light contribution from the nearby galaxy, 
which is almost directly west of the afterglow.
\par
In order to determine the properties of the nearby galaxy 
(hereafter referred to as J1213.1--2054.8) we used 
\emph{SExtractor} (Bertin \& Arnouts \cite{sex})
to obtain its $VRIJ_sK_s$ total magnitudes (\texttt{mag\_auto}). 
Given that our first epoch
data had a very similar seeing in all filters 
($0\parcsec6$--$0\parcsec7$), it was used to extract
the photometry. The PSF-subtracted images, where the afterglow had 
been removed, were given as input to \emph{SExtractor}. We used the same 
detection image for the various filters in order to measure the
flux in an identical aperture. Our results are given in 
Table~\ref{gal.tab}. Note that the galaxy is very red with 
$R - K = 4.3$.
\par
\begin{table}[!t]
\caption[]{Photometry of the galaxy J1213.1--2054.8, $1\parcsec2$ WNW of 
the optical afterglow. No correction for Galactic
extinction has been applied to the photometry.}
\label{gal.tab}
\centering
\scriptsize
\setlength{\arrayrulewidth}{0.8pt}   
\begin{tabular}{ccccc}
\hline
\hline
\vspace{-2 mm} \\
$V$ & $R$ & $I$ & $J_s$ & $K_s$ \\
\vspace{-2 mm} \\
\hline
\vspace{-2 mm} \\
$23.70 \pm 0.18$ & $22.70 \pm 0.12$ & $21.29 \pm 0.09$ & 
$20.50 \pm 0.11$ & $18.40 \pm 0.07$ \\
\vspace{-2 mm} \\
\hline
\end{tabular}
\end{table}     
\begin{table}[!b]
\caption[]{The result of fitting the $R$-band light curve of the 
optical afterglow with a broken power-law. The different methods 
are described in Sect.~\ref{curve.sec}. We note that
$\chi^2_{\mathrm{d.o.f.}}= \chi^2/
\mathrm{degree~of~freedom}$, is the reduced $\chi^2$ of the fit.}
     \label{lc.tab}
\centering
\setlength{\arrayrulewidth}{0.8pt}   
\begin{tabular}{@{}lccccc@{}}
\hline
\hline
\vspace{-2 mm} \\
 & $\chi^2_{\mathrm{d.o.f.}}$ & d.o.f. & $\alpha_1$ &
$\alpha_2$ & $t_{\mathrm{b}}$ [days] \\
\vspace{-2 mm} \\
\hline
\vspace{-2 mm} \\
(i)   & 6.15 &9& $-0.88 \pm 0.03$ & $-2.87 \pm 0.25$ & $2.26 \pm 0.10$\\
(ii)  & 2.33 &6& $-0.99 \pm 0.02$ & $-3.46 \pm 0.44$ & $2.76 \pm 0.11$\\
(iii) & 1.96 &4& $-0.95 \pm 0.03$ & $-1.72$ (fixed)  & $0.78 \pm 0.01$\\
\vspace{-2 mm} \\
\hline
\end{tabular}
\end{table}
Due to the faintness of the OA at $\Delta t \approx 6.65$\,days, and
to the fact that the seeing was comparable to the separation between
the OA and J1213.1--2054.8, it was not possible to perform simple
PSF photometry on the OA. Instead we proceeded in the following way.
We calculated the OA position relative to three nearby stars in 
the first VLT epoch, where the OA was much brighter than 
J1213.1--2054.8. We then created a PSF-subtracted image 
($\Delta t \approx 6.65$\,days) to remove the majority of the 
J1213.1--2054.8 light. At the position of the OA we finally 
subtracted the PSF by varying its magnitude until the residuals
had been adequately removed.
\subsection{Light curve}
\label{curve.sec}
\begin{figure}
   \centering
   \resizebox{\hsize}{!}{\includegraphics[bb=90 370 542 730,clip]{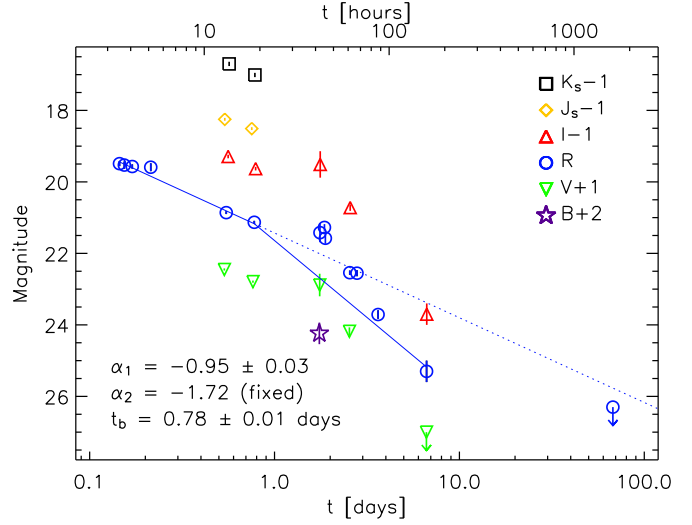}}
      \caption{The optical/NIR light curve of the afterglow based on 
   the measurements
   given in Table~\ref{phot.tab}. The dotted line is an 
   extrapolation of the straight line fit through the early $R$-band 
   data. The power-law indices and time of the break refer to
   the $R$-band fit.}
         \label{lc.fig}
   \end{figure}

Our optical/NIR afterglow light curve is presented in 
Fig.~\ref{lc.fig}. As the $R$-band was better sampled than the 
other filters we analyse it in detail below. We fit the light
curve with a broken power-law (with indices $\alpha_1$ and 
$\alpha_2$ before and after the break, respectively) in three
different ways: \emph{(i)} including all data points in the fit;
\emph{(ii)} excluding the bump around $\Delta t \approx 1.8$\,days, 
assuming it is short-lived and consisting of three points; 
\emph{(iii)} excluding the bump, assuming it is long-lived and
sampled by six points, and fixing $\alpha_2$ at a value obtained
from the SED fitting, where $\alpha_2 = -p$ and $p$ is 
the electron energy power-law index (see Sect.~\ref{ag.sec}). The 
results of the fitting are listed in Table~\ref{lc.tab}.
\par
Due to the bump in the light curve, \emph{(i)} is formally 
rejected with high significance. A better fit is obtained in 
\emph{(ii)} by omitting the three points clustered around 
$\Delta t \approx 1.8$\,days. On the other hand, in this case 
$\Delta \alpha \equiv \alpha_1 - \alpha_2 = 2.47 \pm 0.44$,
a value too high to be consistent with any of the fireball
model predictions (see discussion in Sect.~\ref{ag.sec}).
The best fit is acquired in \emph{(iii)}, indicating there
is an increase in the OA flux above the extra\-polated power-law, 
lasting for approximately 2--3\,days. We note the late-time 
behaviour of the light curve is consistent with the observations 
of Khamitov et al. (\cite{khamitov}) who find $R = 23.5 \pm 0.5$ at 
$\Delta t = 3.43$\,days.
\par
The other filters display a similar initial decay index, with 
$\alpha_V \sim \alpha_I \sim  
\alpha_{Ks} \sim -0.9$ and $\alpha_{Js} \sim -0.7$. 
In addition, the final $I$-band detection clearly implies that the 
light curve has steepened, with a similar
late-time decay index as in the $R$-band. The upper limit in $V$ 
supports the same conclusion for that filter. The bump 
is also present in the $VI$ filters as indicated by the points
obtained with the 1.0-m telescope. The data points covering the bump 
are close enough in time compared to the bump time scale in order
to compute a reasonable estimate of the $V-R$ and $R-I$ colours.
These are consistent with the same colours calculated from two
previous epochs, $\Delta t$\,=\,0.548\,days and
\mbox{$\Delta t = 0.777$\,days} (see Sect.~\ref{sed.sec}). In 
conclusion, the light curve displays an achromatic behaviour
with a bump lasting for a couple of days.
\subsection{Spectral energy distribution of the afterglow}
\label{sed.sec}
\begin{figure}
   \centering
   \resizebox{\hsize}{!}{\includegraphics[bb=75 360 540 730,clip]{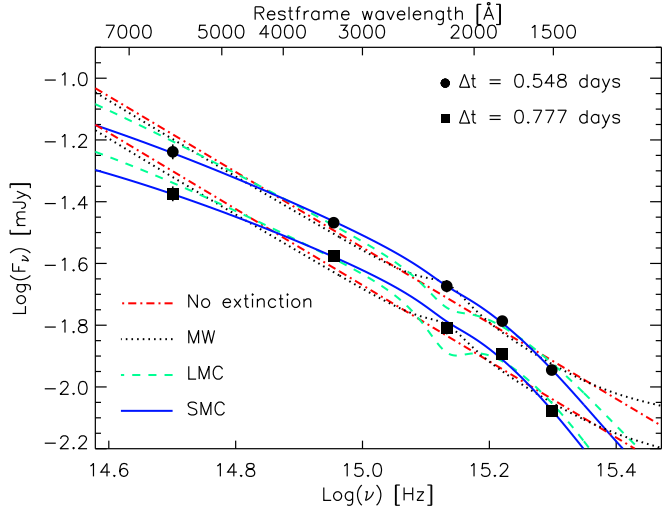}}
      \caption{The spectral energy distribution (SED) of the GRB 030429
      afterglow at $\Delta t = 0.548$\,days (filled circles) and 
      $\Delta t = 0.777$\,days (filled squares). The data points are 
      based on our $K_sJ_sIRV$ measurements. The abscissa displays the 
      frequency/wavelength in the GRB restframe, assuming 
      $z = 2.658$ (see Sect.~\ref{redshift.sec}). At
      both epochs the SED is best fit by an underlying $\beta 
      \simeq -0.36$ power-law with an SMC extinction law with 
      $A_V \simeq 0.34$.}
         \label{sed.fig}
   \end{figure}

Our multiband observations of GRB 030429 allowed the construction 
of the SED at two epochs. We interpolated
the magnitudes to common epochs ($\Delta t$\,=\,0.548\,days and
\mbox{$\Delta t = 0.777$\,days}), using the aforementioned power-law 
indices. We note that the flux from the host galaxy has not been 
subtracted. However, the host is faint enough ($R > 26.3$) as seen 
at $\Delta t \approx 67$\,days, that it should contribute $<$1$\%$ of
the flux at the epochs we are exploring. The SED was constructed as
explained in Fynbo et al. (\cite{sed}). The result is shown in 
Fig.~\ref{sed.fig}, where we have corrected the observed data points 
for foreground (Galactic) extinction using the reddening maps 
of Schlegel et al. (\cite{schlegel}), giving $E(B-V) = 0.062$ at 
that position on the sky.
\par
In order to quantify the effects of extinction we assume that
the intrinsic spectrum of the afterglow is a power-law, and
fit the function $F_{\nu} \propto \nu^{\beta} \times 
10^{-0.4 A_{\nu}}$ to the observed SED, where
$A_{\nu}$ is the extragalactic extinction along the line-of-sight to 
the burst. The dependence of $A_{\nu}$ on $\nu$ has been parameterised
in terms of the restframe $A_V$ following the three extinction laws 
given by Pei (\cite{pei}) for the Milky Way (MW), Large Magellanic 
Cloud (LMC) and Small Magellanic Cloud (SMC). For the assumed 
extinction law, the fit provides $\beta$ and $A_V$ simultaneously
(see e.g. Jensen et al. \cite{brian}; Fynbo et al. \cite{sed};
Holland et al. \cite{holland}; Jakobsson et al. \cite{palli};
Hjorth et al. \cite{jens}; Vreeswijk et al. \cite{paul}). For
comparison purposes we also considered the unextincted case, 
a pure power-law spectrum given by $F_{\nu} \propto \nu^{\beta}$.
\par
The fits are shown in Fig.~\ref{sed.fig}, while the parameters 
of the fits for both epochs are given in Table~\ref{sed.tab}. 
The SMC fit is fully consistent with the data from
both epochs, while other fits are formally rejected with high 
significance. For the redshift of GRB~030429 the interstellar 
extinction bump at 2175\,\AA\ falls into the 
observed $I$-band. As expected by the lack of this absorption bump
in Fig.~\ref{sed.fig}, the MW and LMC extinction are completely 
inconsistent with the data. The absence of the 2175\,\AA\ feature,
often ascribed to graphite grains and ubiquitous in the spectra 
of sight lines through the Galactic diffuse interstellar medium, 
suggests that the GRB host galaxy has lower metallicity and 
dust content than the MW and the LMC. In conclusion, the featureless 
SMC extinction law provides the best fit at both epochs, with 
an average extinction of $A_V = 0.34 \pm 0.04$ and a spectral
index of $\beta = -0.36 \pm 0.12$. This is the most accurate
$A_V$ determination obtained so far for a GRB host galaxy. We note 
that our estimation 
of $A_V$ is not severely limited by the lack of $B$- and $U$-band 
data since the flux observed in those bands
would be attenuated by the Ly$\alpha$ forest. 
At $z = 2.658$ (see Sect.~\ref{redshift.sec}) the 
effective GRB restframe wavelength of these filters is indeed 
located below 1215\,\AA.
\section{Spectroscopy}
\label{spec.sec}
\begin{table}
\caption[]{The results of fitting different extinction laws to our 
GRB~030429 $K_sJ_sIRV$ afterglow observations. We note that
$\chi^2_{\mathrm{d.o.f.}}= \chi^2/
\mathrm{degree~of~freedom}$, is the reduced $\chi^2$ of the fit.}
     \label{sed.tab}
\centering
\setlength{\arrayrulewidth}{0.8pt}   
\begin{tabular}{lcccc@{}}
\hline
\hline
\vspace{-2 mm} \\
Fitting function & $\chi^2_{\mathrm{d.o.f.}}$ & d.o.f. &
$\beta$ & $A_V$  \\
\vspace{-2 mm} \\
\hline
\vspace{-2 mm} \\
\hspace{-2.5 mm}
\emph{$\Delta t = 0.548$~days:} &     &                  & \\  
No extinction & 7.45 & 3 & $-1.22 \pm 0.04$ & 0  \\
MW & 7.84 & 2 & $-1.37 \pm 0.07$ & $<$\,0 \\
LMC & 8.86 & 2 & $-0.70 \pm 0.24$ & $0.30 \pm 0.13$ \\
SMC & 0.14 & 2 & $-0.44 \pm 0.17$ & $0.30 \pm 0.06$ \\
\vspace{-2 mm} \\
\hspace{-2.5 mm}
\emph{$\Delta t = 0.777$~days:} &     &                  & \\  
No extinction & 12.01 & 3 & $-1.22 \pm 0.04$ & 0  \\
MW & 15.50 & 2 & $-1.35 \pm 0.07$ & $<$\,0 \\
LMC & 11.08 & 2 & $-0.31 \pm 0.24$ & $0.52 \pm 0.14$ \\
SMC & 1.17 & 2 & $-0.27 \pm 0.17$ & $0.37 \pm 0.06$ \\
\vspace{-2 mm} \\
\hline
\end{tabular}
\end{table}
\subsection{Reduction and spectral extraction}
The data reduction was performed with standard techniques for bias and
flat-field corrections. The
individual spectra were co-added, and the resulting combined science
frame was sky subtracted in the following way. In the science frame,
regions on both sides of the spectra were filtered along the spatial
axis with a $1 \times 13$ pixels median filter in order to remove cosmic
ray hits. These regions were averaged to produce a 1D sky
spectrum. The mean sky spectrum was expanded to a 2D
spectrum by duplicating the 1D sky spectrum, and this was subtracted 
from the unfiltered science frame.
\begin{figure*}
   \centering
   \includegraphics[width=8.96cm]{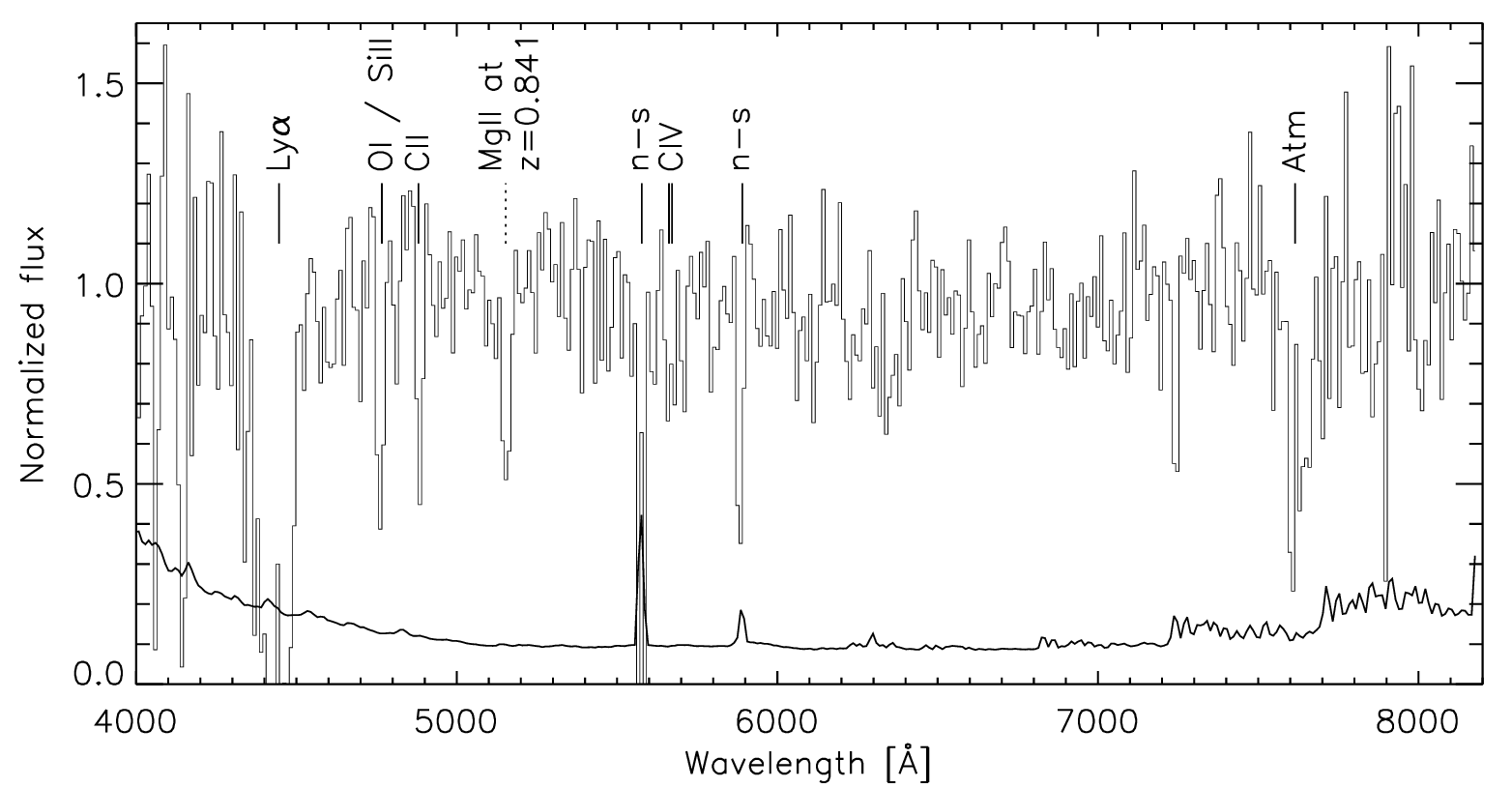}
   \includegraphics[width=8.96cm]{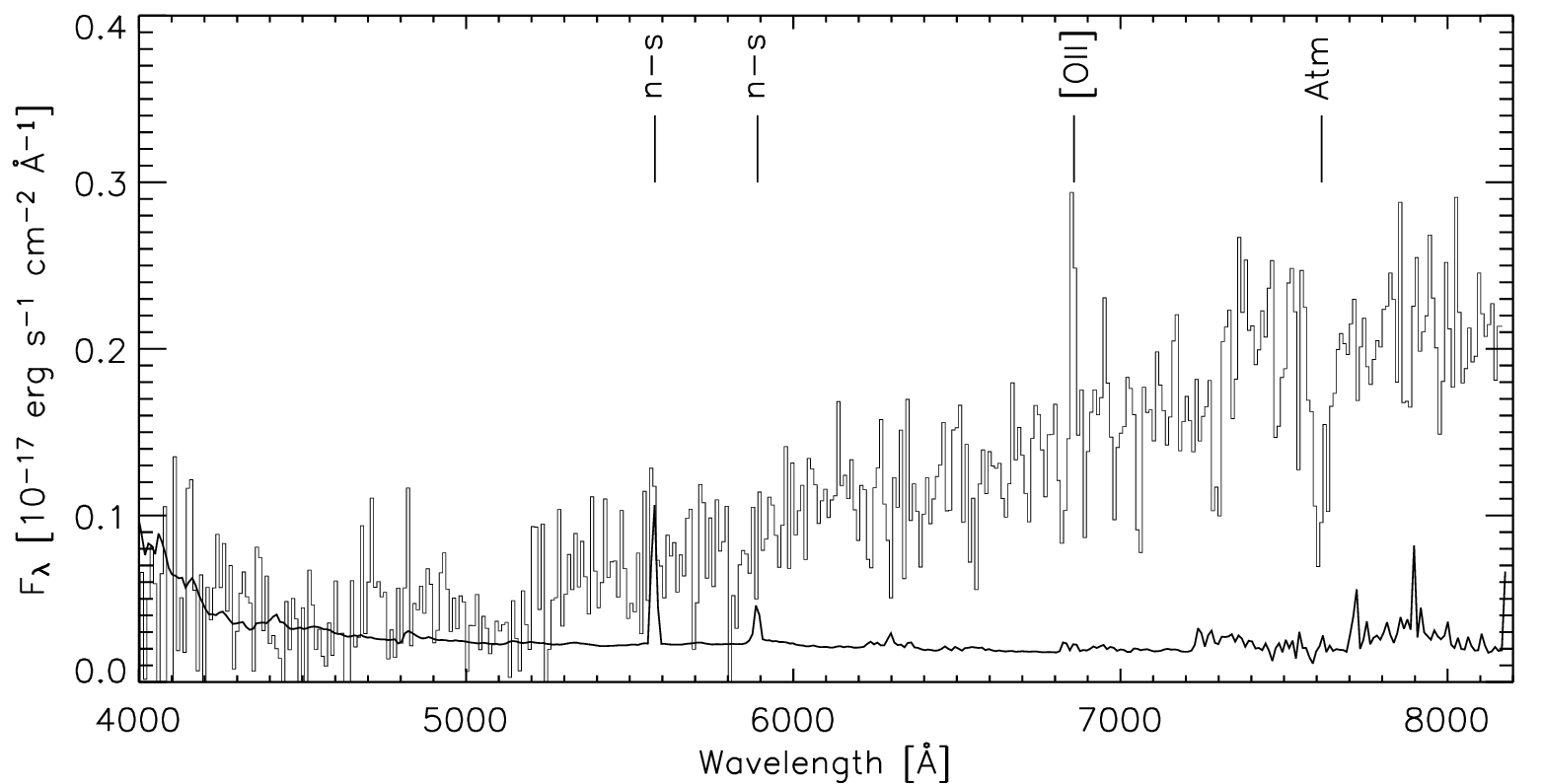}
      \caption{\emph{Left:} the normalized OA spectrum 2.77\,days 
      from the onset of the burst. We detect five
      absorption lines in the spectrum. The 
      identifications are presented in Table~\ref{lines.tab}.
      \emph{Right:} the spectrum of J1213.1--2054.8. We detect one 
      emission line in the spectrum. In both panels the spectra have
      been rebinned to match the size of a resolution element (10\,\AA), 
      and the lower curve is the Poisson error spectrum. Strong 
      residuals from night-sky emission lines are marked with 'n-s'.}
   \label{spec.fig}
\end{figure*}
\par
The spectra were optimally extracted using the code described in
M{\o}ller (\cite{palle}). The output of the code is the
extracted 1D spectra and the 2D residual spectra. The
spectral PSFs of the OA and J1213.1--2054.8 over\-lap in our 2D
spectrum, so in order to separate the two contributions we employed an
iterative procedure: \emph{(i)} extract and remove the OA spectrum; 
\emph{(ii)} extract the 2D spectrum of the neighbouring galaxy and
subtract it from the original spectrum containing both components;
\emph{(iii)} extract the 2D spectrum of the afterglow and subtract it 
from the original spectrum containing both components. After four 
iterations of \emph{(ii)} and \emph{(iii)} a stable solution was found.
\par
The resulting spectra were wavelength-calibrated using the
\texttt{dispcor} task in \emph{IRAF}\footnote{IRAF is distributed by the 
    National Optical Astronomy Observatories,
    which are operated by the Association of Universities for Research
    in Astronomy, Inc., under cooperative agreement with the National
    Science Foundation.}, with an error in the wavelength
    solution of the order of 0.3\,\AA\ (only 3$\%$ of our spectral 
    resolution). Flux calibration was performed using an observation 
    of the standard star LTT7987. Finally, the spectra were rescaled to 
    match the $V$-band magnitude at the same epoch. In Fig.~\ref{spec.fig} 
   we show the spectra of both the OA and J1213.1--2054.8.
\subsection{Redshift}
\label{redshift.sec}
\begin{table}
\caption[]{Features detected in both the OA and J1213.1--2054.8
  spectra. The columns show the line identification, the restframe
  wavelength, the observed wavelength, the restframe
  equivalent width (negative for absorption lines) and the inferred 
  redshift of the lines, respectively.}
\label{lines.tab}
\centering
\setlength{\arrayrulewidth}{0.8pt}   
\begin{tabular}{lccrc}
\hline
\hline
\vspace{-2 mm} \\
Line & $\lambda_{\mathrm{rest}}$ & $\lambda_{\mathrm{obs}}$ 
& $W_{\mathrm{rest}}$ \hspace*{0.3cm} & $z$ \\
     & [\AA] & [\AA] & [\AA] \hspace*{0.35cm} & \\ 
\vspace{-2 mm} \\
\hline
\vspace{-2 mm} \\
\hspace{-3 mm}
\emph{Afterglow:} \\
Ly$\alpha$                & ---      & ---      &$42.1 \pm 2.9$ & --- \\
\ion{O}{i} / \ion{Si}{ii} & $1303.3$ & $4764.8$ & $2.2 \pm 0.4$ & $2.6559$ \\
\ion{C}{ii}               & $1334.5$ & $4885.7$ & $2.5 \pm 0.5$ & $2.6610$ \\
\ion{C}{iv} (doublet)     & ---      & ---      & $1.4 \pm 0.4$ & --- \\
\ion{Mg}{ii} (doublet)    & ---      & ---      & $3.3 \pm 0.4$ & $0.8418$ \\
\hspace{-3 mm}
\emph{J1213.1--2054.8:} \\
$[$\ion{O}{ii}]   & $3727.1$ & $6858.0$ & $-6.8 \pm 0.8$ & $0.8400$ \\
\vspace{-2 mm} \\
\hline
\end{tabular}
\end{table}     
We identify three absorption lines at $>$$5\sigma$ in the spectrum of 
the OA. We use two of them, namely the narrow lines \ion{O}{i} / 
\ion{Si}{ii} and \ion{C}{ii}, to calculate the redshift of the GRB, 
placing it at $z = 2.658 \pm 0.004$. Armed with this redshift we 
searched for additional lines at $>$$3\sigma$. At the expected position 
of the \ion{C}{iv} doublet we find a likely candidate, although at our
spectral resolution we are unable to resolve it. In the
spectrum of J1213.1--2054.8 we detect one emission line, which we
tentatively identify as [\ion{O}{ii}] at a redshift of $z = 0.84$. 
Checking this redshift against the OA spectrum, we identify an absorption 
line at the location of the \ion{Mg}{ii} doublet. The redshift of this 
blended \ion{Mg}{ii} line is based on a minimum $\chi^2$ fitting of
the expected line profile. We conclude that the redshift of the 
nearby galaxy is $z = 0.841 \pm 0.001$. The detected lines and their 
characteristics are listed in Table~\ref{lines.tab}. These results rule 
out J1213.1--2054.8 as being the GRB~030429 host galaxy.
\subsection{The damped Ly$\alpha$ absorption line}
   \begin{figure}
   \centering
   \resizebox{\hsize}{!}{\includegraphics[bb=53 268 540 540,clip]{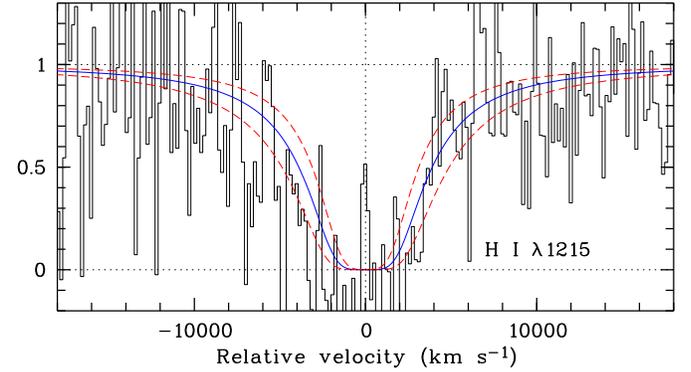}}
      \caption{The normalized OA spectrum centred on the Ly$\alpha$
      absorption line. A neutral hydrogen column density 
      fit to the damped Ly$\alpha$ line is shown with a solid line, while 
      1$\sigma$ errors are shown with dashed lines. }
         \label{dla.fig}
   \end{figure}
Absorption systems with neutral hydrogen column density larger than
$2 \times 10^{20}$\,cm$^{-2}$ are referred to as damped Ly$\alpha$ 
absorbers (DLAs) as they display strong Ly$\alpha$ damping 
wings even in low/medium 
resolution spectra. These are normally detected in QSO spectra
but so far four GRB afterglows have shown evidence for a high column
density DLA system (see e.g. Fig.~4 in Vreeswijk et al. 
\cite{paul}). Almost all of the neutral gas in the \mbox{high-$z$} 
Universe 
is located in the DLAs and they represent a major fraction of the 
baryon reservoir out of which stars at the present epoch formed
(e.g. Storrie-Lombardi \& Wolfe \cite{storrie} and references
therein).
\begin{table*}
\caption[]{Computation of the closure
relation, $|\alpha_1| + b |\beta| + c$, for two afterglow
models. A favourable model will have a value of zero for the 
closure relation. The ISM and wind models are for isotropic
expansion in a homogeneous and wind-stratified environment,
respectively. The electron energy power-law index, $p$, is obtained
in two ways, from $\beta$ and $\alpha_1$. In addition,
$\Delta \alpha$ is calculated from the resulting $p$ in both cases.
The numbers in the table are based on case \emph{(iii)} in 
Table~\ref{lc.tab}.}
     \label{closure.tab}
\centering
\setlength{\arrayrulewidth}{0.8pt}   
\begin{tabular}{ccrrccc}
\hline
\hline
\vspace{-2 mm} \\
Model & $\nu_{\mathrm{c}}$ & Closure \hspace*{2mm} &
$p(\beta)$ \hspace*{11mm} & $\Delta \alpha$ & $p(\alpha_1)$ & 
$\Delta \alpha$ \\
\vspace{-2 mm} \\
\hline
\vspace{-2 mm} \\

ISM & $\nu_{\mathrm{c}} > \nu_{\mathrm{o}}$ &
$0.41 \pm 0.18$ & $1 - 2 \beta = 1.72 \pm 0.24$ & $(p+3)/4 = 1.18 \pm 0.06$ &
$(4|\alpha_1|+3)/3 = 2.27 \pm 0.04$ & $1.32 \pm 0.01$ \\

ISM & $\nu_{\mathrm{c}} < \nu_{\mathrm{o}}$ &
--- \hspace*{4.7mm}& $-2 \beta = 0.72 \pm 0.24$ & --- &
--- & --- \\

Wind & $\nu_{\mathrm{c}} > \nu_{\mathrm{o}}$ &
$-0.09 \pm 0.18$ & $1 - 2 \beta = 1.72 \pm 0.24$ & $(p+1)/4 = 0.68 \pm 0.06$ &
$(4|\alpha_1|+1)/3 = 1.60 \pm 0.04$ & $0.65 \pm 0.01$\\

Wind & $\nu_{\mathrm{c}} < \nu_{\mathrm{o}}$ &
--- \hspace*{4.7mm} & $-2 \beta = 0.72 \pm 0.24$ & --- &
--- & --- \\

\vspace{-2 mm} \\
\hline
\end{tabular}
\end{table*}
\par
In Fig.~\ref{dla.fig} we plot the normalized OA spectral region 
around Ly$\alpha$. Overplotted is a fit to the strong Ly$\alpha$
absorption line yielding $\log N(\ion{H}{i}) = 21.6 \pm 0.2$, 
within the range of previously known GRB-DLAs 
$(21.2 \leq \log N(\ion{H}{i}) \leq 21.9)$. The redshift deduced in the
previous subsection, $z = 2.658$, was imposed on the fit.
Assuming $R_V = 2.91$ and using $A_V = 0.34 \pm 0.04$ from the best 
fit extinction law (SMC), the ratio between the neutral hydrogen 
column density and the reddening in the GRB~030429 host is 
$N(\ion{H}{i}) / E(B-V) = (34 \pm 16) \times 
10^{21}$\,cm$^{-2}$\,mag$^{-1}$. We note that this is a lower limit 
as J1213.1--2054.8 could contribute to the measured $A_V$. This 
value is consistent with that of the SMC, \mbox{$(44 \pm 7) \times 
10^{21}$\,cm$^{-2}$\,mag$^{-1}$}, and similar to those found
for other GRB absorbers (see Table~3 in Hjorth et al. \cite{jens}).
\par
There is an indication of Ly$\alpha$ emission in the centre of the
trough, but it is not statistically significant ($\lesssim$\,2$\sigma$).
The flux of the line, if real, is approximately
$2.5 \times 10^{-18}$\,erg\,s$^{-1}$\,cm$^{-2}$. 
In our assumed cosmology this corresponds to a luminosity of 
$1.5 \times 10^{41}$\,erg\,s$^{-1}$. We can
use this result to estimate the star-formation
rate (SFR), assuming the conversion from Ly$\alpha$ luminosity to the
SFR of $10^{42}$\,erg\,s$^{-1} = 1\,M_{\odot}\,\textrm{yr}^{-1}$
(Kennicutt \cite{kennicutt}; Cowie \& Hu \cite{hu}). The Ly$\alpha$ 
SFR in the host of GRB~030429 is thus 
$\sim$0.15\,$M_{\odot}\,\textrm{yr}^{-1}$. If confirmed
this will strengthen the conclusion of Fynbo et al. (\cite{fy}) that 
Ly$\alpha$ emission is much more frequent among GRB host galaxies 
than among the Lyman-Break galaxies at similar redshifts and that 
there could be a low metallicity preference for GRBs (see also 
Woosley \& MacFadyen \cite{woosley}).
\section{Discussion}
\label{dis.sec}
\subsection{Comparison with afterglow models}
\label{ag.sec}
The nature of the ambient medium in which the GRB originated can
be probed with the parameters $\alpha_1$, $\alpha_2$ and $\beta$.
We consider two afterglow models: an isotropic expansion into a 
homogeneous medium (Sari et al. \cite{ISM}), and an
isotropic expansion into a wind-stratified medium (Chevalier \& Li 
\cite{wind}). We use the closure relation introduced by Price et al. 
(\cite{price02}) in order to differentiate between the models. 
In our notation $|\alpha_1| + b |\beta| + c = 0$, where the values 
of $b$ and $c$ depend on the location of the cooling frequency,
$\nu_{\mathrm{c}}$, relative to the optical/NIR bands,
$\nu_{\mathrm{o}}$, at the epoch of the observations. 
\par
In Table~\ref{closure.tab} we have used the average value of the
spectral index, $\beta=-0.36\pm 0.12$, to derive the 
electron energy power-law index,
resulting in $p = -2 \beta = 0.72 \pm 0.24$ if 
$\nu_{\mathrm{c}} < \nu_{\mathrm{o}}$ and $p = 1 - 2\beta = 1.72 \pm 
0.24$ if $\nu_{\mathrm{c}} > \nu_{\mathrm{o}}$. The former value is 
not considered relevant in afterglow models, and the accurate analytical 
spectral and temporal indices have not been derived in the literature 
for this case.
\par
The ISM model is not favoured by the closure relation. We have used
the results from case \emph{(iii)} (see Sect.~\ref{curve.sec}),
but note that \emph{(i)} and \emph{(ii)} give a similar result as
$\alpha_1$ does not vary significantly between all three cases.
\par
Each spectral or temporal power-law index relates to a certain value
of the electron energy power-law index, $p$, and the correct model
should result in a similar $p$ for all indices. Assuming that
the fireball model is valid, this simple fact excludes \emph{(i)}
and \emph{(ii)}. In these cases $p(\beta) \neq p(\alpha_2) \neq
p(\alpha_1)$, and the predicted $\Delta \alpha$ is in addition
approximately twice as small as the observed one.
\par
In conclusion, only the wind ($\nu_{\mathrm{c}} > \nu_{\mathrm{o}}$)
model produces a closure relation consistent with zero. We have 
fixed $\alpha_2 = - p(\beta)$,
case \emph{(iii)}, which results in an observed $\Delta
\alpha = 0.77 \pm 0.03$. This is marginally consistent with the
predicted values listed in Table~\ref{closure.tab}.
\par
The value of $p$ as estimated from $\beta$, $p(\beta)=1.72 \pm 0.24$, 
is marginally consistent with $p=2$. We derived the closure relations 
for the flat spectrum case, $1<p<2$ (Dai \& Cheng \cite{dai2}; 
Bhattacharya \cite{basi}), and found that although formally consistent 
with the observationally determined value of $p(\beta)$, none of the 
closures were consistent with zero. In addition, the predicted 
$\Delta \alpha$ values were quite far from that observed. This suggest 
that the $p \gtrsim 2$ case is a better representation of the data, although 
not perfect. This may be due to insufficient sampling of the light cure 
and its overall structure (e.g. bumps).
\par
Accepting the above scenario, it is clear from Fig.~\ref{lc.fig}
that there is a bright long-lived achromatic bump above the late-time 
light curve fit. Due to the sparse data sampling we cannot 
constrain the properties of this bump in detail, 
but it lasts for 2--3\,days and seems to
peak at 1.5 magnitudes above the fit. Similar undulations have been
observed previously in light curves associated with GRBs:
GRB~970508 (e.g. Galama et al. \cite{gal}; Sokolov et al. \cite{sok}), 
GRB~000301C (e.g. Sagar et al. \cite{sagar}; Garnavich et al. 
\cite{garna}; Masetti et al. \cite{mas}; Berger et al. \cite{berg}; 
Jensen et al. \cite{brian};
Rhoads \& Fruchter \cite{rhoads}; Dai \& Lu \cite{dai}; Panaitescu 
\cite{pan}; Gaudi et al. \cite{gaudi}), GRB~000911 (Ramirez-Ruiz et 
al. \cite{enrico}), GRB~011211 (e.g. Holland et al. \cite{holl}; 
Jakobsson et al. \cite{palli2}),
GRB~021004 (e.g. Lazzati et al \cite{lazzati}; Bersier et al. 
\cite{bersier}; Heyl \& Perna \cite{heyl}; Holland et al. 
\cite{holland}; Mirabal et al. \cite{mirabal}; Fox et al. \cite{fox}) 
and GRB~030329 (e.g. Lipkin et al. \cite{lipkin} and references 
therein).
\par
 \begin{figure}
\centering
\resizebox{\hsize}{!}{\includegraphics[bb=86 374 534 698,clip]{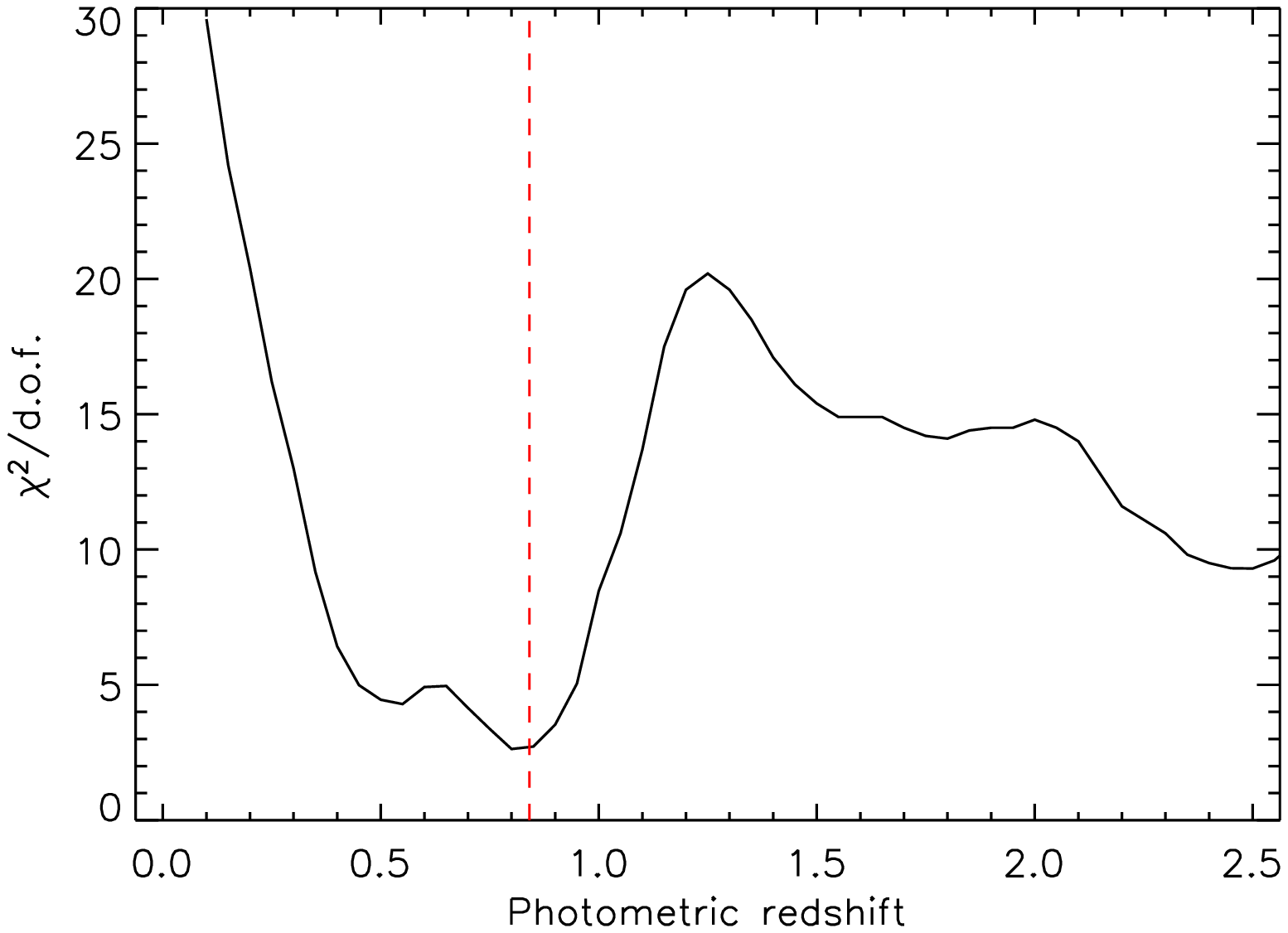}}
\resizebox{\hsize}{!}{\includegraphics[bb=88 364 534 710,clip]{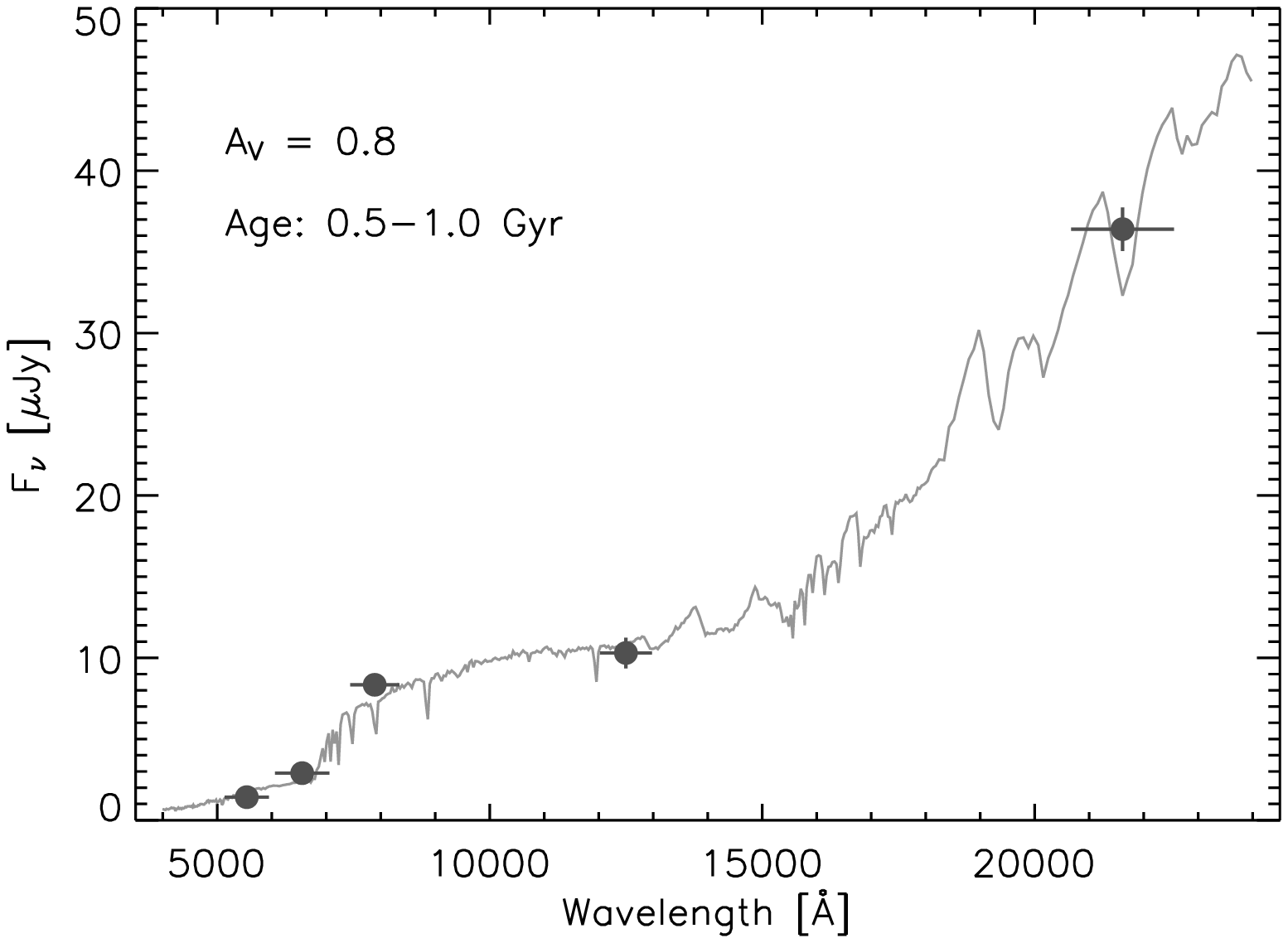}}
   \caption{\emph{Top:} the reduced $\chi^2$ as a function of the
   photometric redshift of the nearby galaxy, obtained from \emph{hyperz}. 
   The best fit is obtained for $z = 0.82$. The 
   spectroscopic redshift of 0.841 is indicated by the vertical dashed 
   line. \emph{Bottom:} the best fit template galaxy spectrum
   along with the observed $VRIJ_sK_s$ 
   SED based on the measurements given in Table~\ref{gal.tab} 
   (filled circles).} 
     \label{hyperz.fig}
 \end{figure}
Deviations from the common smooth power-law decay have been
explained by a variable external density (e.g. Wang \& Loeb 
\cite{wl}), refreshed shocks from the inner engine (e.g. 
Rees \& M\'{e}sz\'{a}ros \cite{rm98}), a non-uniform jet
structure (e.g. M\'esz\'aros et al. \cite{mes98}) or
a microlensing event (e.g. Garnavich et al. \cite{garna}). 
Simultaneous multi-wavelength observations can
help to determine which of the above scenarios is responsible for
OA light curve variations (e.g. Nakar \& Piran \cite{nakar};
Nakar et al. \cite{nakar2}; Jakobsson et al. 
\cite{palli2}). Armed only with a sparse optical/NIR data set for
GRB 030429, we are unable to distinguish between these
possibilities.
\subsection{The intervening absorption galaxy}
In order to gain more information about J1213.1--2054.8 we used the 
photometric redshift code \emph{hyperz} (Bolzonella et al. 
\cite{hyperz}). In the fitting process the programme uses 
standard $\chi^2$ minimisation, i.e. computing and minimising the 
deviations between the observed SED (see Table~\ref{gal.tab}) of an 
object and template SEDs compared within the same photometric
system. For J1213.1--2054.8 the best fit is obtained for a photometric
redshift of $z = 0.82 \pm 0.02$, in 
excellent agreement with the spectroscopic redshift of 0.841 that we
previously determined (upper panel of Fig.~\ref{hyperz.fig}).
\par
The best fit template spectrum is displayed in  
Fig.~\ref{hyperz.fig} (lower panel). It corresponds to a 
fairly young stellar population, \mbox{age between} 
0.5--1.0\,Gyr, with a significant extinction of
\mbox{$A_V = 0.8$}. In addition, in J1213.1--2054.8 we have
detected [\ion{O}{ii}] in emission (Table~\ref{lines.tab}). 
All evidence is thus 
consistent with a star-forming galaxy. We calculate
its absolute restframe $R$-band magnitude by transforming the 
observed $J$-band magnitude (Table~\ref{gal.tab}) to AB magnitudes
(Fukugita et al. \cite{ABmags}), applying the distance modulus
and the $2.5 \log (1+z)$ term (see e.g. Appendix A in van Dokkum \& 
Franx \cite{dokkum}). This results in $M_R(\mathrm{AB}) = -21.6$,
a value located on the bright end of the luminosity function.
\par
The galaxy is located $1\parcsec2$ from the line-of-sight to the 
OA, which corresponds to an impact parameter of $D = 9$\,kpc at
$z = 0.841$ in our adopted cosmology. $D$ is frequently interpreted 
as the lower limit
of the radii of suspected counterparts of QSO absorption-selected 
galaxies. Guillemin \& Bergeron (\cite{guillemin}) reported that 
galaxy counterparts of \ion{Mg}{ii} absorbers with equivalent 
widths larger than 0.6\,\AA\ and $0.7<z<1.3$, all had $D > 50$\,kpc. 
That result was consistent with the conclusions of Bergeron \& Boiss\'e 
(\cite{bergeron}), who examined a sample of $z \sim 0.5$ \ion{Mg}{ii}
absorbers (see also Steidel \cite{steidel}). 
\par
In every case, the possibility exists that the real absorber is 
very faint and/or hidden inside the glare of the QSO, thus 
overestimating the suspected counterpart radius. Indeed, more than 
half of present-day galaxies reside in groups (Eke et al. \cite{eke}), 
lending support to this hypothesis. The fact that many OAs are high-$z$ 
(Fig.~\ref{z.fig}) and transient phenomena has made it possible to
search for faint/nearby absorbers. The OAs fade away completely 
within a few months, so it is only a matter of obtaining long
exposures in order to get deep limits. GRB~030429 is the second burst
in which a \ion{Mg}{ii} absorber with a small $D$ has been identified 
with imaging and spectroscopy. The GRB~020405 sight line also
revealed a spectroscopically confirmed intervening \ion{Mg}{ii} 
absorber with an impact para\-meter of only 13\,kpc (Masetti et 
al. \cite{masetti}). 
\begin{table}
\caption[]{GRBs whose spectra display an \ion{Mg}{ii} absorption 
system besides the one produced in the host galaxy. Here, $D$ is the 
impact para\-meter, the projected distance between the OA and the 
candidate \ion{Mg}{ii} absorber. The impact parameter is given in kpc 
where the absorber is spectroscopically confirmed.}
     \label{MgII.tab}
\centering
\setlength{\arrayrulewidth}{0.8pt}   
\begin{tabular}{ccrc}
\hline
\hline
\vspace{-2 mm} \\
GRB & $z_{\mathrm{GRB}}$ & $z_{\mathrm{abs}}$ \hspace*{0.42cm} & $D$ \\
\vspace{-2 mm} \\
\hline
\vspace{-2 mm} \\
970508 & 0.835 & 0.768        & $\sim$$5\parcsec5$ \\
991216 & 1.022 & 0.770, 0.803 & $\sim$$0\parcsec5$, $\sim$$2\parcsec0$ \\
010222 & 1.477 & 1.156, 0.927 & $\sim$$2\parcsec0$, $\sim$$4\parcsec0$\\
020405 & 0.691 & 0.472        & 13\,kpc (2\parcsec0) \\
021004 & 2.335 & 1.381, 1.604 & $\sim$$16\arcsec$ \\
030226 & 1.986 & 1.043        & $\sim$$2\parcsec5$ \\
030429 & 2.658 & 0.841        & 9\,kpc (1\parcsec2) \\
\vspace{-2 mm} \\
\hline
\end{tabular}
\end{table}
\par
In Table~\ref{MgII.tab} we list all GRBs that have an \ion{Mg}{ii} 
absorption system along the line of sight in addition to that
originating from the host galaxy. We have omitted 
bursts whose spectra display an \ion{Mg}{ii} absorption that most 
likely arises in a system physically interacting with the host 
(GRB~000926: Castro et al. \cite{castro}; GRB~020813: Barth et al. 
\cite{barth}). Of the seven 
bursts, two (GRB~020405 and GRB~030429) have a 
spectroscopically confirmed counterpart with a very small impact
parameter of $D \sim 10$\,kpc. The host of GRB~991216 is located
close to at least three nearby galaxies which are likely candidates
for the two \ion{Mg}{ii} absorption systems (Vreeswijk 
\cite{paul_phd}). Should that be the case, their impact parameters
would correspond to only 4--15\,kpc. The situation is similar for
GRB~010222, where at least four nearby galaxies surround its host
(e.g. Mirabal et al. \cite{mirabal02}; Frail et al. \cite{frail};
Galama et al. \cite{gal03}). Two galaxies are located $2\parcsec5$
away from GRB~030226, corresponding to 20\,kpc at $z = 1.043$ 
(e.g. Price et al. \cite{price03}; Klose et al. \cite{klose}). 
A possible galaxy counterpart to
the \ion{Mg}{ii} system in GRB~970508 is located $5\parcsec5$ away
from the host, corresponding to 40\,kpc at $z = 0.786$ (e.g. Metzger 
et al. \cite{metzger}; Djorgovski et al. \cite{djorg}; Pian et al. 
\cite{pian}; Sokolov et al. \cite{sok2};
Fructher et al. \cite{andy}). For GRB~021004 a possible galaxy candidate 
at an impact parameter of $16\arcsec$, corresponding to 135\,kpc at
$z = 1.381$, has been observed in narrow-band imaging (Vreeswijk 
et al. \cite{paul03}). The authors point out that a
deeper study should be performed on the GRB~021004 field before
a faint absorber at a low impact parameter can be ruled out.
\par
A large fraction of possible OAs 
appears to have a faint absorber at small $D$. These results 
imply that some previous 
QSO absorption-selected galaxy counterpart identifications may be
incorrect, with the real absorber being a nearby faint galaxy.
This suggestion is supported by the observations of 
Falomo et al. (\cite{falomo})
and Watson et al. (\cite{darach}), who find a likely counterpart to
an \ion{Mg}{ii} absorption system in a BL Lac object at only
$D = 11$\,kpc. In addition, Ellison et al. (\cite{ella2}) 
analysed 27 intervening absorption systems in the 
spectra of a triply imaged QSO. They find that the most likely
absorber coherence scale is 3\,kpc, more than one order of magnitude 
smaller than the dimensions of \ion{Mg}{ii} absorbers deduced from 
the impact parameters. 
\subsection{Strong gravitational lensing?}
With an angular separation of only $1\parcsec2$, J1213.1--2054.8 
could affect the appearance of the OA via strong gravitational 
lensing (e.g. Narayan \& Bartelmann \cite{lens}; Grossman \& Nowak
\cite{grblenses}). For
that to occur the Einstein radius must be at least
$1\parcsec2$, which corresponds to a lens mass of 
$\gtrsim$$5.4 \times 10^{11}\,M_{\odot}$. Using 
\mbox{$M_R(\mathrm{AB}) = -21.6$} gives a mass-to-light ratio 
of $\gtrsim$$15\,\Upsilon_{\odot}$ (where $\Upsilon_{\odot} \equiv
M_{\odot}/L_{\odot}$). At \mbox{$z = 0.841$} we are only 
probing the innermost 9\,kpc of the galaxy. The aforementioned mass 
is roughly six times larger than for Milky Way-like galaxies within 
a similar radius. In addition, the 
critical surface mass density for this configuration is 
0.43\,g\,cm$^{-2}$, a factor of three to four times that in normal 
galaxy lenses (see discussion regarding GRB~990123 in Andersen et 
al. \cite{michael}). 
\par
We also utilised the \emph{lensmodel} package (Keeton \cite{keeton})
to model J1213.1-2054.8 as a singular isothermal ellipsoid model with 
a flat rotation curve. Using our $R$-band image from $\Delta t = 
67.6$\,days we constrained its ellipticity, $e \approx 0.16$, and 
position angle, $\phi \approx -49^{\circ}$. Knowing the distance
between the lens and the OA, we varied the lens mass until the 
\texttt{findsrc} output resulted in a multiple image system.
This occurred when the mass inside 9\,kpc 
reached $\gtrsim$$3 \times 10^{11}\,M_{\odot}$, a value comparable 
to the one obtained above. In conclusion, the three independent 
methods discussed here virtually rule out that GRB~030429 was 
multiply imaged (with a time delay of roughly a month) by the nearby 
galaxy.
\section{Conclusions}
\label{con.sec}
GRB~030429 occurred in a faint ($R > 26.3$) galaxy at a redshift of
$z = 2.658 \pm 0.004$. The derived neutral hydrogen column density
is $\log N(\ion{H}{i}) = 21.6 \pm 0.2$, while the restframe
reddening, obtained from the optical/NIR SED, is accurately measured
to be $A_V = 0.34 \pm 0.04$. We conclude that the high value of the
ratio between column density and optical extinction is readily 
explained by a metal-poor environment, similar to that of 
the SMC. This is fully consistent with the favoured SMC-like 
extinction law obtained from the SED fit.
\par
A nearby galaxy with a separation of only $1\parcsec2$ is
ruled out as the host, with $z = 0.841 \pm 0.001$. This is the second
time a galaxy adjacent to a GRB OA is found to be responsible 
for a \ion{Mg}{ii} absorption system in the OA spectrum. In both
cases the impact parameter was small, $D \sim 10$\,kpc. At least three
additional bursts have galaxy counterparts with similar small impact 
parameters that presumably account for their \ion{Mg}{ii} absorption 
systems. Thus, at least five out of seven GRB OAs displaying \ion{Mg}{ii} 
in absorption (see Table~\ref{MgII.tab}), have a nearby galaxy with
$D \sim 10$--20\,kpc. This strongly indicates that previous 
identifications of many QSO absorption-selected galaxy counterparts
are possibly incorrect. The remaining OA galaxy counterparts should be 
spectroscopically confirmed before a firmer conclusion can be drawn.
\par
Within the framework of the afterglow synchrotron model 
(Sari et al. \cite{ISM}), the GRB~030429 early-time decay 
index, $\alpha_1$, and spectral index, $\beta$, indicate that a jet 
($\nu_{\mathrm{c}} > \nu_{\mathrm{o}}$) expanding into a wind
medium provides the most likely scenario for the data. The late-time 
decay index, $\alpha_2$,
has to produce an electron energy power-law index, $p$, equivalent
to that calculated from $\beta$. This suggests that a bright 
long-lived bump is present in the GRB 030429 light curve. We 
emphasise that densely sampled GRB light curves are of the utmost
importance in order to observe these variations; without the 
$\Delta t \approx 1.8$\,days GRB~030429 observations it would
have been difficult to detect the flux increment above the 
interpolated power-law.
\par
\vspace*{0.4cm}
\emph{\hspace*{-0.5cm}Note added in manuscript.}---After the submission of 
this paper, GRB~030429 was 
classified\footnote{\texttt{http://space.mit.edu/HETE/Bursts/Data/}} as 
an X-ray flash (XRF). Together with the unambiguous OA redshift 
determination (see Sect.~\ref{redshift.sec}), this makes it the first 
XRF with a spectroscopically confirmed absorption redshift. We note
that within the uncertainties the burst is consistent with being an
X-ray rich GRB. The most probable reason for this burst being classified
as an XRF is a combination of its fairly low intrinsic $E_{\mathrm{peak}}$,
and the relatively high redshift of the burst (resulting in an observed 
$E_{\mathrm{peak}} \approx 35$\,keV).

\begin{acknowledgements}
     We thank the anonymous referee for a very positive and constructive
     report. We are grateful to Roland Vanderspek for providing us with
     various GRB~030429 properties derived from analyses of HETE data.
     PJ, GB and EHG acknowledge support from a special 
     grant from the Icelandic Research Council. JPUF and KP acknowledge 
     support from the Carlsberg foundation. This work was supported by the 
     Danish Natural Science Research Council (SNF). The authors acknowledge 
     benefits from collaboration within the EU FP5 Research Training 
     Network "Gamma-Ray Bursts: An Enigma and a Tool".
\end{acknowledgements}

\end{document}